\documentclass[pra,twocolumn,showpacs,superscriptaddress,10pt]{revtex4-1}
\usepackage[colorlinks,citecolor=blue,linkcolor=red,dvipdfm]{hyperref}
\usepackage{amsmath,amssymb,graphicx}
\usepackage{times}
\usepackage{float}

\begin{document}

\title{Light gap bullets in defocusing media with optical lattices}

\author{Zhiming Chen}
\affiliation{State Key Laboratory of Transient Optics and Photonics, Xi'an
Institute of Optics and Precision Mechanics of Chinese Academy of Sciences, Xi'an 710119, China}
\affiliation{School of Science, East China University of Technology, Nanchang 330013, China}

\author{Jianhua Zeng}
\email{\underline{zengjh@opt.ac.cn}}
\affiliation{State Key Laboratory of Transient Optics and Photonics, Xi'an
Institute of Optics and Precision Mechanics of Chinese Academy of Sciences, Xi'an 710119, China}
\affiliation{University of Chinese Academy of Sciences, Beijing 100049, China}
\date{\today}

\begin{abstract}
Searching for three-dimensional spatiotemporal solitons (also known as light/optical bullets) has recently attracted keen theoretical and experimental interests in nonlinear physics. Currently, optical lattices of diverse kinds have been introduced to the stabilization of light bullets, while the investigation for the light bullets of gap type---nonlinear localized modes within the finite gap of the underlying linear Bloch spectrum---is lacking. Herein, we address the formation and stabilization properties of such light gap bullets in periodic media with defocusing nonlinearity, theoretically and in numerical ways. The periodic media are based on two-dimensional periodic standing waves created in a coherent three-level atomic system which is driven to the regime of electromagnetically induced transparency, which in principle can also be replaced by photonic crystals in optics or optical lattices in ground-state ultracold atoms system. The temporal dispersion term is tuned to normal (positive) group velocity dispersion so that to launch the light gap bullets under self-repulsive nonlinearity; two types of such light gap bullets constructed as 3D gap solitons and vortices with topological charge $m=1$ within the first finite gap are reported and found to be robustly stable in the existence domains. On account of the light bullets were previously limited to the semi-infinite gap of periodic media and continuous nonlinear physical systems, the light gap bullets reported here thus supplement the missing type of three-dimensional spatiotemporal localized modes in periodic media which exhibit finite band gaps.
\end{abstract}

\maketitle

$\emph{Introduction.}$---It is a challenging issue to create stable three-dimensional (3D) localized modes (alias solitons) owning to the inherent supercritical wave collapse triggered by attractive cubic (Kerr) nonlinearity (critical wave collapse also exists in two-dimensional (2D) settings)~\cite{OL-RMP,NL-RMP,RJP-review1,NRP,PR-BEC,2D-3Dbook,wave-collapse,focusing-collapse,discrete-collapse}.  To against such high-dimensional wave collapses in the soliton research field in diverse branches of science, it is therefore necessary to introduce extra physical effects, including synthetic periodic potentials~\cite{DarkGS,CP,PT3D,APR,GW,EIT-OL2d,opt1,opt2,opt3,opt6,opt7}, saturable absorber~\cite{absorber1,absorber2}, optical cavity~\cite{cavity1,cavity2,cavity3}, semiconductor active~\cite{semiconductor-active} or quadratic nonlinear media~\cite{Quadratic}, waveguide and fiber arrays~\cite{waveguide1,waveguide2,waveguide3,waveguide4}, materials with nonlocal~\cite{nonlocal1,nonlocal2} or competing (focusing) cubic and (defocusing) quintic nonlinearities~\cite{CQ1,CQ2}, linear spin-orbit coupling~\cite{SOC}, etc.

The creation of 3D localized modes including light bullets (spatiotemporal solitons) has recently attracted a great deal of attention in diverse optical systems and Bose-Einstein condensates (BECs) from both theoretical predictions~\cite{Bullet-Bessel0,Rev05,Bullet-Bessel,liquidCrystal,HuangGX4,HuangGX5,HuangGX6,HuangGX8,HuangGX9,soliton-laser,bullet-Fan,moire-bullet} and experimental observations~\cite{waveguide1,waveguide2,waveguide3}. Particularly, theoretical predictions demonstrated the storage and retrieval of $(3 + 1)$-dimensional low power (weak-light) bullets and vortices in ultracold $^{87}$Rb atomic gases under the electromagnetically induced transparency (EIT) regime~\cite{HuangGX4}, such idea was then extended to ultracold Rydberg atomic gases which exhibit both the local and nonlocal nonlinearities~\cite{HuangGX6,HuangGX9}. Kartarshov {et al.} addressed the creation and properties of stable 3D topological solitons (light bullets) in focusing Kerr nonlinear media with parity-time symmetric optical lattices~\cite{HuangGX5}. Recently, Kartarshov predicted intriguing stability regime of light bullets in photonic moir\'{e} lattices~\cite{moire-bullet}. All these studies about light bullets in the absence/presence of periodic potentials are, however, restricted to focusing nonlinear media; while the generation, stabilization, and properties of light bullets in the form of 3D localized gap modes (which we call light gap bullets), residing within the finite gaps of the associated linear Bloch-wave spectrum, in periodic defocusing nonlinear media, to the best of our knowledge, remain unexplored.

Here, we address, theoretically and numerically, the existence and stability properties of 3D spatiotemporal localized gap states (alias light gap bullets) in defocusing atomic media working in EIT condition trapped by 2D optical lattice. Two classes of topologically localized gap states in the form of fundamental gap solitons (zero vorticity) and gap vortices with vorticity (topological charge) $m=1$ are reported, with the finding that they are robustly stable within the first finite band gap, and unstable close to the both edges of Bloch bands. Despite light bullets have been investigated intensively in focusing media with periodic potentials, they are exclusively confined to the semi-infinite gap (but not the finite gap) of the associated linear Bloch band. Thus, our study reports the first gap-type light bullets in periodic optical media with repulsive nonlinearity, opening a new way for exploring high-dimensional stable topological gap modes such as light bullets in 2D periodic media.

\begin{figure}
\centering
\includegraphics[scale=0.41]{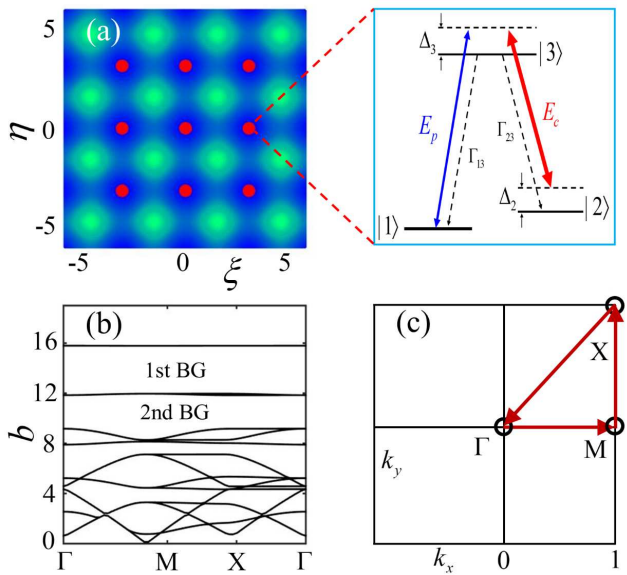}
\caption{Coherent atom ensembles trapped by 2D OLs and the associated linear Bloch spectrum. (a) Contour plot of a 2D square OL with depth $c_0=10$, green shading corresponds to lattice potential maxima,  red points denote the three-level $\Lambda$-type atoms (whose exciation scheme is depicted in top right) trapped in the lattice  minima. (Right-top) $^{87}$Rb atoms tuned to D1-line transition~\cite{Steck} under EIT mode: A weak probe field $\textbf{E}_p$  and strong continuous-wave control field $\textbf{E}_c$ couple, respectively, the transitions $|1\rangle \leftrightarrow |3\rangle$ and $|2\rangle \leftrightarrow |3\rangle$; $\Gamma_{13}$ and $\Gamma_{23}$ being spontaneous emission decay rates of $|3\rangle \rightarrow |1\rangle$ and $|3\rangle \rightarrow |2\rangle$ transitions, and the associated detunings are $\Delta_2$ and $\Delta_3$. (b) The linear band-gap spectrum with 1st/2nd BG represents the first/second band-gap. (c) The first reduced Brillouin zone of high symmetry points in reciprocal lattice space.
}\label{fig1}
\end{figure}

$\emph{Theoretical model.}$---Propagation dynamics of 3D spatiotemporal solitons in periodic optical media or electromagnetically induced optical lattices is governed by the dimensionless nonlinear Schr\"{o}dinger equation for complex amplitude, $\psi(\xi,\eta,\tau) $, along propagation distance $s$ (see Supplementary Material):
\begin{equation}
i\frac{\partial \psi}{\partial s}=-\frac{1}{2}(\frac{\partial^2\psi}{\partial \xi^2}+\frac{\partial^2\psi }{\partial \eta^2})+\frac{\beta}{2}
\frac{\partial^2\psi}{\partial \tau^2}+g|\psi|^2\psi+V_{\textrm{OL}}\psi,
\label{GP}
\end{equation}
where $\xi$ and $\eta$ are coordinates in the transverse plane, and $\tau$ is time, the 2D optical lattice yields $V_{\textrm{OL}}(\xi,\eta)=-c_0(\sin^2 \xi+\sin^2 \eta$), we set strength $c_0=10$ for discussion, and $g$ being the generalized nonlinear Kerr coefficient. As depicted in Fig. \ref{fig1}(a), for a three-level $\Lambda$-type coherent atomic ensemble where EIT turns on, such 2D optical lattice could be formed by pairs of counter-propagating far-detuned laser (Stark) fields. The governing equation [Eq. (\ref{GP})] is derived from the Maxwell-Bloch equations by using the multiple scales method under electric-dipole and rotating-wave approximations (more details are provided in Supplementary Material). The linear Bloch spectrum of such lattice is displayed in Fig. \ref{fig1}(b), showing a wide first finite gap; the corresponding reduced Brillouin zone follows the general way of square lattices [Fig. \ref{fig1}(c)].

The remaining parameter $\beta$ in Eq. (\ref{GP}) denotes the dispersion coefficient, which is fixed in the anomalous group-velocity dispersion (GVD) region as  $\beta<0$ for the generation of fundamental bright soliton solutions; such rule was always obeyed by entire past studies of spatiotemporal solitons in the presence/absence of periodic potentials, to our best knowledge. Corresponds what with this is the 3D spatiotemporal solitons supported by 2D complex lattices were reported so far merely in the self-focusing (attractive, $g<0$) nonlinearity regime. In particular, the periodic potentials have shown incomparable band-gap control engineering within where new localized states called localized gap modes like gap solitons and vortices were and still are warmly explored in both experimental and theoretical sides~\cite{DarkGS,CP,PT3D,APR,GW,EIT-OL2d,opt1,opt2,opt3,opt6,opt7}. It is therefore a pronounced imagine is to reveal the formation of 3D spatiotemporal solitons in finite gaps of the 2D optical lattices and how and what their stabilization and dynamics configurations would look like.

The conjecture physically sounds, as evidenced by two fundamental realities: (i) the localized gap modes exist under self-defocusing (repulsive, $g>0$) nonlinear regime, which can be easily achieved in experimental aspect; (ii) to construct gap-type 3D spatiotemporal solitons (light gap bullets) under $g>0$, the anomalous GVD ($\beta<0$) should be reversed to the normal GVD regime ($\beta>0$), which could be alternatively selected by the wavelength of light (in nonlinear optics) or positive effective mass via band-gap modulation (in BECs illumated by pairs of counterpropagating laser beams). Accordingly, it is technically safe and of great interest to present theoretical understanding of 3D light gap bullets, which are fundamentally new  spatiotemporal localized structures. This work aims to that mission.

The stationary gap-mode solution $\psi$ at propagation constant $b$ is expressed as $\psi=\phi e^{ib\tau}$, then Eq. (\ref{GP}) reduces to the stationary equation
\begin{equation}
b \phi=\frac{1}{2}(\frac{\partial^2\phi}{\partial \xi^2}+\frac{\partial^2\phi }{\partial \eta^2})-\frac{\beta}{2}
\frac{\partial^2\phi}{\partial \tau^2}-|\phi|^2\phi-V_{\textrm{OL}}\phi.
\label{station}
\end{equation}
In the following, self-defocusing nonlinearity is discussed and thus we set $g\equiv1$. To show dependency $P(b)$ below, we define the soliton power as $P=\int\int\int_{-\infty}^{+\infty}|\phi(\xi,\eta, \tau)|^2d\xi d\eta d\tau$.

Before presenting numerical results we describe the adopted numerical methods. Specifically, the stationary light-gap-bullet solutions are sought from Eq.~(\ref{station}) by means of modified squared-operator iteration method~\cite{MSOM},  and their stability is checked in direct perturbed simulations [Eq.~(\ref{GP})] via fourth-order Runge-Kutta method.

\begin{figure*}
\centering
\includegraphics[scale=0.7]{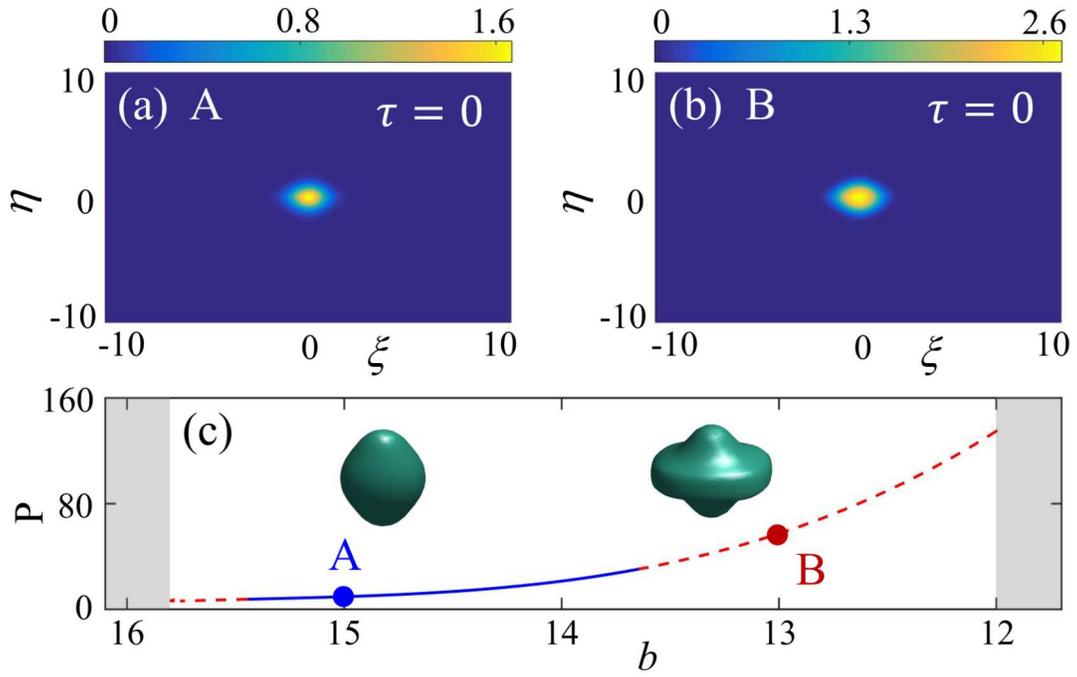}
\caption{Shapes and existence domain of light gap bullets created in the first band-gap of the 2D OLs. Contour plots of (a) stable and (b) unstable gap bullets with parameters: (a) $b=15$, $P=9.564$; (b) $b=13$, $P=57.34$. (c) Stability (solid) and instability (dashed) regions of light gap bullets shown as energy (soliton power $P$) of light gap bullets versus propagation constant ($b$), displayed are also for isosurfaces of stable and unstable ones, whose contour plots are shown in (a) and (b) respectively.}\label{fig2}
\end{figure*}

\begin{figure*}
\centering
\includegraphics[scale=0.7]{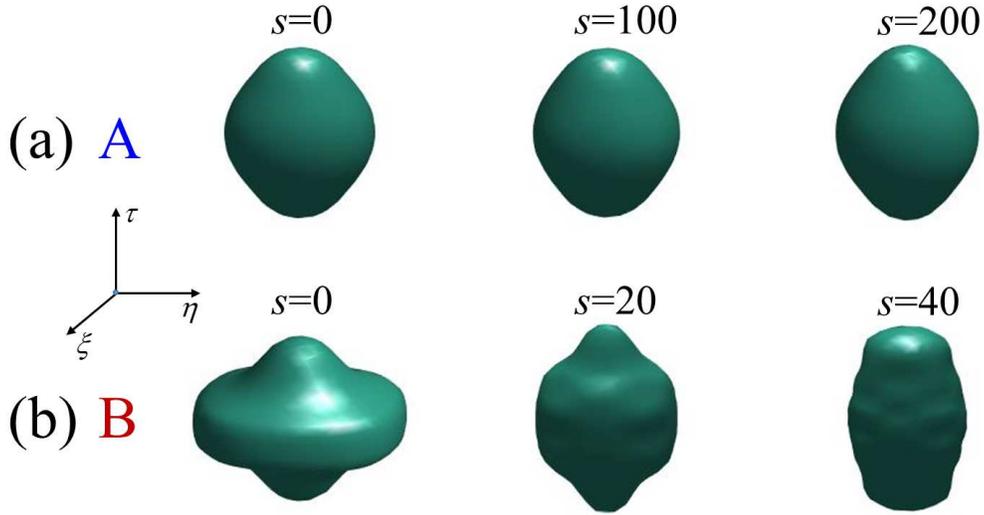}
\caption{Perturbed propagation dynamics of stable (a) and unstable (b) light gap bullets [corresponding to marked points in Fig.~\ref{fig1}(c)] along propagation distance $s$.}\label{fig3}
\end{figure*}

\begin{figure*}
\centering
\includegraphics[scale=0.7]{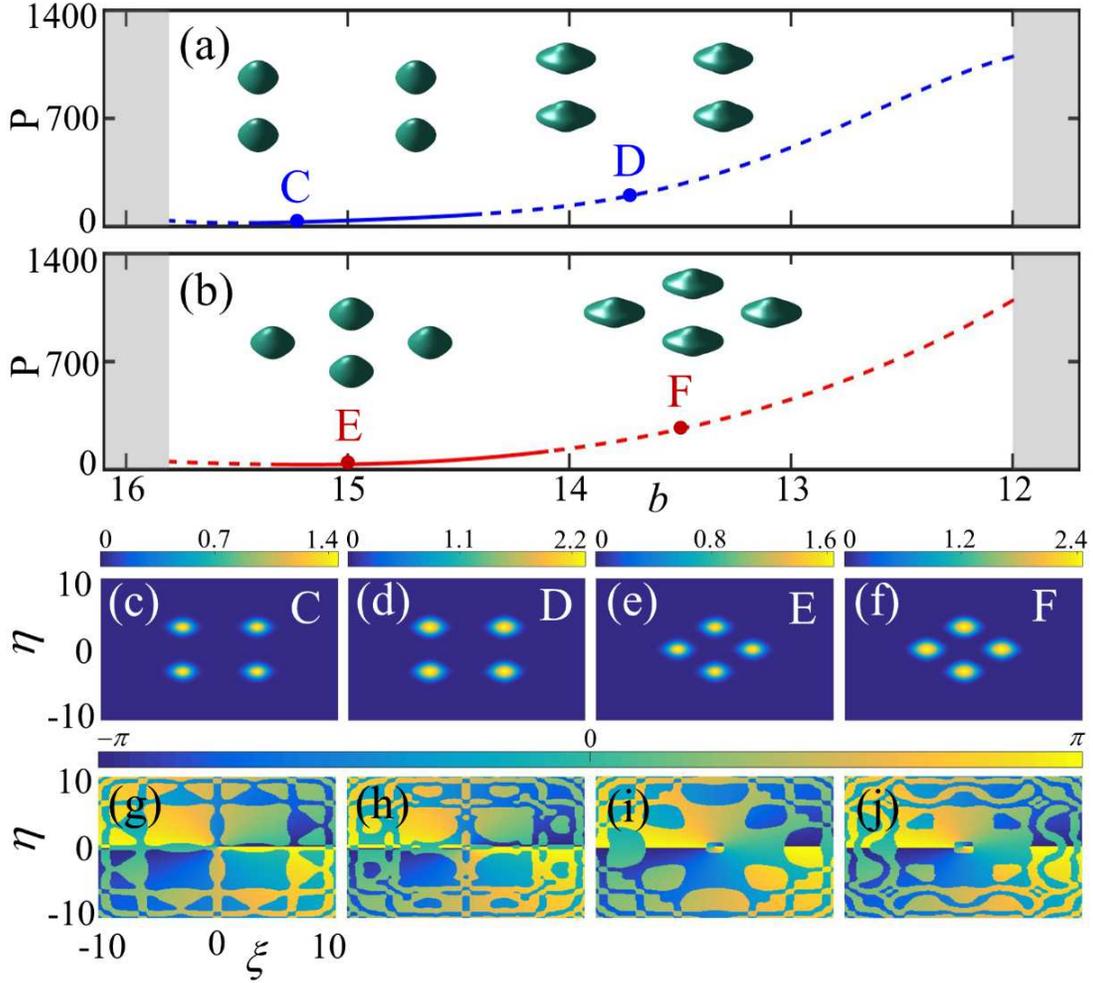}
\caption{Shapes, phase structures, and existence domains of light gap vortices with topological charge $m=1$ in the first band-gap of the 2D OLs. Energy (soliton power $P$) of light gap vortices versus $b$: stability (solid) and instability (dashed) regions of off-site (a) and on-site (b) light gap bullets, whose representative isosurfaces marked as points ($\mathbf{C}$, $\mathbf{D}$, $\mathbf{E}$, $\mathbf{F}$) are also depicted in the panels and the contour plots and the associated phase distributions are displayed, respectively, in the third and bottom lines. }\label{fig4}
\end{figure*}

\begin{figure*}
\centering
\includegraphics[scale=0.7]{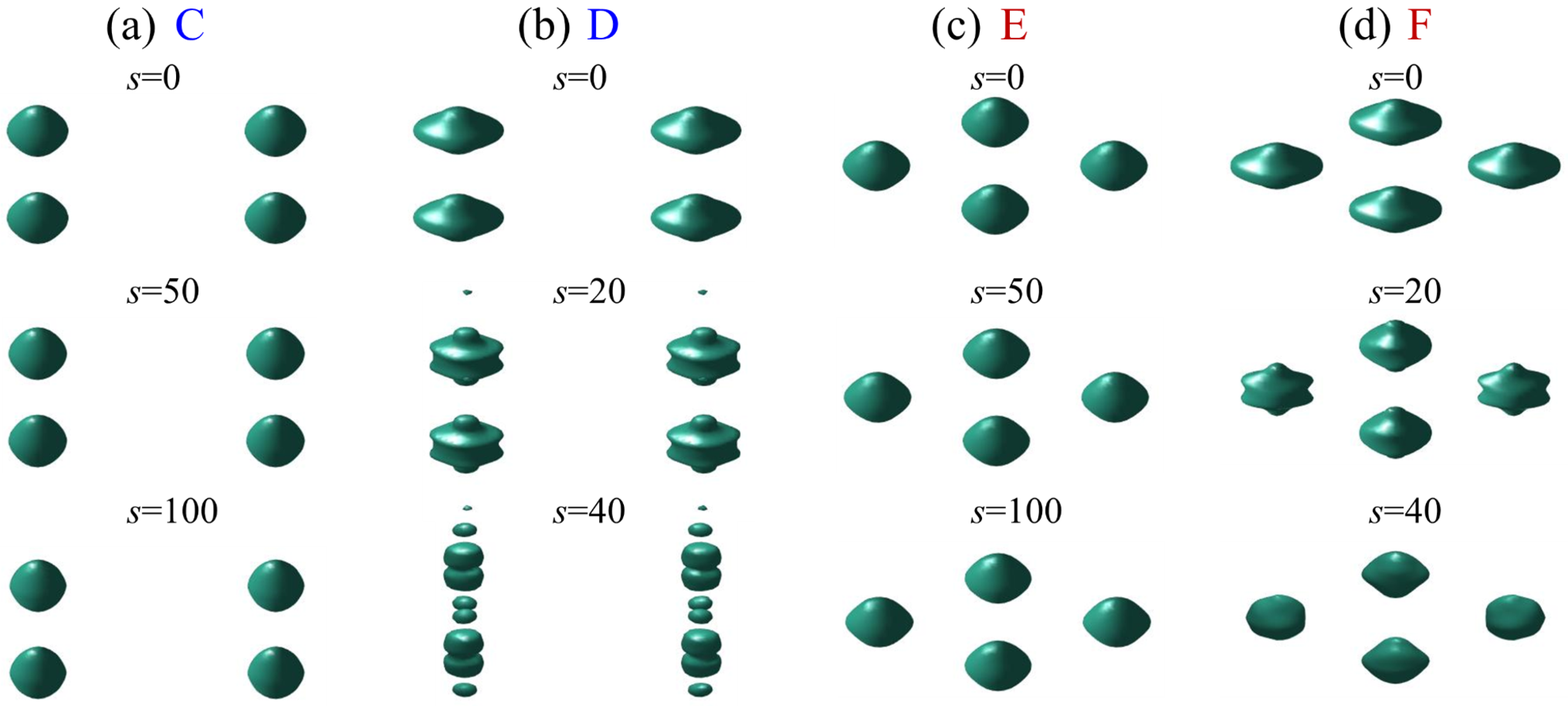}
\caption{Perturbed propagation dynamics of stable (a, c) and unstable (b, d) off-site (a, b) and on-site (c, d) light gap bullets [corresponding to marked points in Figs.~\ref{fig4}(a) and~\ref{fig4}(b)] along propagation distance $s$.}\label{fig5}
\end{figure*}



$\emph{Light gap bullets.}$---Gap solitons are among the fundamental gap-type mode in optical lattices, which including temporal term constitutes the first and simplest kind of light gap bullets. Spatial configurations of a stable and unstable examples of which populated at the first band-gap are displayed (as contour plots at $\tau=0$), respectively, in  Figs. \ref{fig2}(a) and \ref{fig2}(b) , the corresponding isosurface plots (considering time axis) are given in Fig. \ref{fig2}(c) where also shows the $P(b)$ curve for such light gap bullets. It is readily seen from the last panel that they are stable only in the midst of first finite gap and, particularly, such stability region is narrower than that for their 2D counterparts (2D configurations without time term, i.e., $\beta=0$). We stress that the temporal derivative (for 3D spatiotemporal mode only)---which is not confined in 2D spatial lattice directions---does not reinforce, but, conversely, weaken the stability of light gap bullets could explain this.

Displayed in Figs. \ref{fig3}(a) and \ref{fig3}(b) are the perturbed propagation dynamics of the both stable and unstable light gap bullets under study. According to the evolutional process, we can observe that good coherence conserves for the former, while for the latter, its shape in the time axis ($\tau$) and two spatial coordinates ($\xi$, $\eta$) changes drastically. The unstable dynamics therefore poses a significant challenge to fully control the generation of stable light gap bullets within a broad stability region.

Therefore, this section of the numerical evidence facts reports the first stable gap-type light bullets in periodic potentials under self-defocusing nonlinearity and normal GVD, this finding demonstrates the above-mentioned, reasonable conjuncture, which we firstly conceive based on logical deduction and physical principle.

$\emph{Light gap vortices.}$---The second kind of light gap bullets is light gap vortices, which are considered as 3D vortex gap solitons with topological charge $m=1$, and which are usually excited as hollow quadruple localized-wave structures consisted of four identical compositions (fundamental gap solitons), each of them inhabits the lattice potential minimum and is connected with a $\pi/2$ phase displacement as a nested topological solitons. Depending on their spatial positions centered on and centered not on the axes $(\xi=0$ or $\eta=0)$, such quadruple-soliton modes carrying a topological dislocation can be called on-axis and off-axis light gap vortices. Typical isosurface profiles of them are displayed in Figs. \ref{fig4}(a) and \ref{fig4}(b) , portrayed in where are also the corresponding dependence $P(b)$ for both topological modes. It is seen that the energy of a 3D spatiotemporal vortex soliton (of both modes) is about four times larger than that of a single light gap bullet, and remarkably, the stability domain of the former shrinks a little bit. Consistent with 3D vortex gap solitons in other settings, the on-axis 3D light gap bullets have a wider stability region than their off-axis counterparts [compare Figs. \ref{fig4}(a) and \ref{fig4}(b)]. The contour plots of the typical stable and unstable off-axis light gap bullets and on-axis ones are shown in turn in Figs. \ref{fig4}(c) and \ref{fig4}(d), and Figs. \ref{fig4}(e) and \ref{fig4}(f), and the homologous phase distributions are respectively shown in Figs. \ref{fig4}(g) and \ref{fig4}(h), and Figs. \ref{fig4}(i) and \ref{fig4}(j). It should be noted that the phase structures of the light gap vortices feature a continuous and gradual change from $-\pi$ to  $\pi$,  which clearly indicate how the hollow quadruple structures are connected to each other. The light gap vortices fundamentally provide an alternative manifestation for in-depth research of soliton molecules, complex soliton modes carrying with diverse topological charge and phase singularity, and soliton-soliton interactions.

Figs. \ref{fig5}(a) and \ref{fig5}(b) displayed respectively the perturbed propagation dynamics of stable and unstable off-axis light gap bullets, showing clearly that the generation of stable 3D spatiotemporal vortex gap soliton which sustain its shape and energy during propagation for the former, and for the latter, the unstable one evolves into uncertain transformed shape that can no longer be called soliton. Similar dynamics are also disclosed for on-axis light gap vortices, according to Figs. \ref{fig5}(c) and \ref{fig5}(d).

$\emph{Experimental feasibility.}$---It is worth noting that the 2D periodic potentials of different configurations could be readily realized as photonic crystals and lattices using the current state-of-the-art technologies (which have become mature) in optics context, and as optical lattices in ultracold atoms. Besides, by choosing suitable laser wavelength one can achieve the normal GVD regime in optical materials, such regime can also be realized by positive effective mass in BECs~\cite{NL-RMP,NRP,PR-BEC,OL-RMP}. These two  physically realizable contexts are well described by our governing model [Eq. (\ref{GP})], and accordingly, the new localized modes as light gap bullets and vortical ones are observable in both platforms.

$\emph{Conclusion.}$---We have revealed the existence and properties of light gap bullets and their vortical ones in defocusing media with 2D optical lattices induced in three-level atomic ensembles working under EIT regime. Based upon numerical simulations and theoretical analysis, we gave the (in)stability domains of both 3D spatiotemporal localized modes within the first finite gap. Since the light bullets have not yet been explored inside the finite gaps of periodic potentials, our results thus fill a gap of the missing kind of 3D spatiotemporal localized-wave structures formed inside the gaps, which are observable in nonlinear optics or ultracold atoms loaded onto optical lattices, suggesting an approach to the creation of new type of 3D spatiotemporal solitons in the self-defocusing media with low-dimensional periodic potentials in terms of scientific interest, paving the way to enhance greatly the efficiency of light-matter interactions and to multifunctional optical information-processing in practical applications.


\medskip
\textbf{Acknowledgements} \par 
This work was supported by the National Natural Science Foundation of China (NSFC) (Nos. 12074423, 12264002, and 12264003), Young Scholar of Chinese Academy
 of Sciences in western China (XAB2021YN18), and Jiangxi Provincial Natural Science Foundation (20202BABL211013).

\medskip
\textbf{Author contributions}\par
Z.C. carried out the analytical and numerical calculations, and wrote the manuscript. J.Z. conceived the idea, conducted the calculation and revised the manuscript.

\medskip
\textbf{Conflict of interest}\par
The authors declare no competing interests.

\medskip
\textbf{Supplementary information} The online version contains supplementary
material available at https://doi.org/

{}
\end{document}


\title{Supplementary Material for: Light gap bullets in defocusing media with optical lattices}

\author{Zhiming Chen}
\affiliation{State Key Laboratory of Transient Optics and Photonics, Xi'an
Institute of Optics and Precision Mechanics of Chinese Academy of Sciences, Xi'an 710119, China}
\affiliation{School of Science, East China University of Technology, Nanchang 330013, China}

\author{Jianhua Zeng}
\email{\underline{zengjh@opt.ac.cn}}
\affiliation{State Key Laboratory of Transient Optics and Photonics, Xi'an
Institute of Optics and Precision Mechanics of Chinese Academy of Sciences, Xi'an 710119, China}
\affiliation{University of Chinese Academy of Sciences, Beijing 100049, China}
\begin{abstract}
This supplementary documents provide further details on the explicit expressions of Maxwell-Bloch equations under electromagnetically induced transparency (EIT) and deriving the nonlinear envelope equation of the probe field that governs the fundament of generating light gap bullets in the main text. In Sec.~\ref{Sec:1}, we depict the physical model under consider and give the explicit expressions of the underlying Maxwell-Bloch equations. We present the elaborate derivation of the nonlinear envelope equation under standard multiple scales method developed for EIT system in Sec.~\ref{Sec:2}.
\end{abstract}
\date{\today}
\maketitle

\captionsetup[figure]{labelfont={bf},labelformat={default},labelsep=period,name={Fig.}}

\section{Theoretical model and expressions of Maxwell-Bloch equations}{\label{Sec:1}}

The physical model under consider is a lifetime-broadened three-level $\Lambda$-type coherent atomic ensemble interacting resonantly with two laser fields (i.e.,  the pulsed probe field $\mathbf{E}_p$ and the strong continuous-wave control field $\mathbf{E}_c$ ) and working under the condition of electromagnetically induced transparency (EIT)~\cite{EIT-OL2d} [See Fig. 1(a) in the main text]. Particularly, the atoms are assumed under an ultralow temperature to eliminate the Doppler broadening effect and illuminated by pairs of counter-propagating far-detuned laser (Stark) fields that form two-dimendional (2D) optical lattices to creating light gap bullets. In addition, the weak probe field with center angular frequency $\omega_p$ couples ground state $|1\rangle$ to excite state $|3\rangle$, and the strong control field with center angular frequency $\omega_c$ drives the transitions from metastable state $|2\rangle$ to excite state $|3\rangle$. $\Gamma_{13}$ and $\Gamma_{23}$ are the spontaneous emission decay rates of transitions $|3\rangle \rightarrow |1\rangle$ and $|3\rangle \rightarrow |2\rangle$, respectively. The detunings $\Delta_3=\omega_{p}-\omega_{31}$ and $\Delta_2=\omega_{p}-\omega_{c}-\omega_{21}$ represent respectively one- and two-photon detunings. Note that here $\omega_{jl}=(E_j-E_l)/\hbar$ with $E_j$ being the eigen energy of state $|j\rangle$.

For simplification, here we set both the probe and control laser fields propagating along the $z$ direction, thus the electric-field vector in our model can be written as $\textbf{E}=\hat{\mathbf{e}}_{p}{\cal E}_{p} \exp[{i(k_{p} z-\omega_{p} t)]}+\hat{\mathbf{e}}_{c}{\cal E}_{c} \exp{[i(k_{c} z-\omega_{c} t)]}+\textrm{c.c.}$. Here $\hat{\mathbf{e}}_{p}$ ($\hat{\mathbf{e}}_{c}$) is the unit vector of the probe (control) field with the envelope ${\cal E}_p$ (${\cal E}_c$), and $k_{p}=\omega_{p}/c$ ($k_c=\omega_c/c$) is the wavenumber of the probe (control) field before entering the atomic gas. Furthermore, to stabilize light gap bullets in 2D optical lattices, we assume two pairs of counter-propagating Stark laser fields are applied to the three-level $\Lambda$-type coherent atomic ensemble [See Fig. 1(a) in the main text]. The form of far-detuned Stark fields is considered as
$\textbf{E}_{\textrm{Stark}}(x,y,t) = \sqrt{2}[\hat{\textbf{e}}_{s1} E_{s1}(x)+\hat{\textbf{e}}_{s2} E_{s2}(y)]\cos(\omega_st)$,
where $\hat{\textbf{e}}_{s1}$ ($\hat{\textbf{e}}_{s2}$), $E_{s1}$ ($E_{s2}$), and $\omega_s$ are the unit polarization vector, field amplitude, and angular frequency, respectively. It should be noted that $\hat{\textbf{e}}_{s1}$ is assumptively perpendicular to $\hat{\textbf{e}}_{s2}$. The Stark fields can cause a small but space-dependent energy shift for the level $|j\rangle$ in the $x$ and $y$ directions, which have the form $\Delta E_{j,\textrm{Stark}}=-\frac{\alpha_j}{2}[|E_s(x)|^2+|E_s(y)|^2]$ with $\alpha_j$ being the scalar polarizability of the level $|j\rangle$.

%
Under the methods of electric-dipole and rotating-wave approximations, the Hamiltonian of the system in the interaction picture is
$$\hat{\cal H}_{\rm int}=-\sum_{j=1}^{3}\hbar\Delta_j^\prime|j\rangle\langle j|-\hbar\,\left[\Omega_c|3\rangle\langle2|+\Omega_{p}|3\rangle\langle1|+{\rm H.c.}\right],\eqno{(\textrm{S1})}$$ with $\Delta_j^\prime=\Delta_j+\frac{1}{2}\alpha_{j1}[|E_{s1}(x)|^2+|E_{s2}(y)|^2]$ and $\alpha_{j1}=(\alpha_{j}-\alpha_{1})/\hbar$.
Here half Rabi frequencies of the control and probe fields are respectively defined as $\Omega_c=(\mathbf{p}_{23}\cdot{\hat{\mathbf {e}}}_c){\cal E}_c/\hbar$ and $\Omega_{p}=(\mathbf{p}_{13}\cdot{\hat{\mathbf {e}}}_{p}){\cal E}_{p}/\hbar$ with $\textbf{p}_{jl}$ the electric dipole matrix element related to the transition from $|j\rangle$ to $|l\rangle$.
Thus the equation of motion for density matrix $\sigma$ in the interaction picture is given by
$$i\frac{\partial}{\partial
t}\sigma_{11}-i\Gamma_{13}\sigma_{33}+\Omega_p^{\ast}\sigma_{31}-\Omega_p\sigma_{31}^{\ast}=0, \eqno{(\textrm{S2a})}$$
$$i\frac{\partial}{\partial
t}\sigma_{22}-i\Gamma_{23}\sigma_{33}+\Omega_{c}^{\ast}\sigma_{32}-\Omega_{c}\sigma_{32}^{\ast}=0, \eqno{(\textrm{S2b})}$$
$$i\frac{\partial}{\partial
t}\sigma_{33}+i(\Gamma_{13}+\Gamma_{23})\sigma_{33}-\Omega_p^{\ast}\sigma_{31}+\Omega_p\sigma_{31}^{\ast}\-\Omega_c^{\ast}\sigma_{32}+\Omega_c\sigma_{32}^{\ast}=0, \eqno{(\textrm{S2c})}$$
$$\left(i\frac{\partial}{\partial
t}+d_{21}\right)\sigma_{21}-\Omega_p\sigma_{32}^{\ast}+\Omega_c^{\ast}\sigma_{31}=0,\eqno{(\textrm{S2d})}$$
$$\left(i\frac{\partial}{\partial
t}+d_{31}\right)\sigma_{31}-\Omega_p(\sigma_{33}-\sigma_{11})+\Omega_c\sigma_{21}=0,\eqno{(\textrm{S2e})}$$
$$\left(i\frac{\partial}{\partial
t}+d_{32}\right)\sigma_{32}-\Omega_c(\sigma_{33}-\sigma_{22})+\Omega_p\sigma_{21}^{\ast}=0,\eqno{(\textrm{S2f})}$$
where $d_{jl}=\Delta_{j}^\prime-\Delta_{l}^\prime+i\gamma_{jl}$. Dephasing rates are defined as $\gamma_{jl}=(\Gamma_j+\Gamma_l)/2+\gamma_{jl}^{\rm col}$ with $\Gamma_j=\sum_{E_i<E_j}\Gamma_{ij}$  being the spontaneous emission rate from the state $|j\rangle$ to all lower energy states $|i\rangle$ and $\gamma_{jl}^{\rm col}$  being the dephasing rate reflecting the loss of phase coherence between $|j\rangle$ and $|l\rangle$.
The Maxwell equation for the probe-field Rabi frequency $\Omega_{p}$ can be achieved by taking the method of slowly varying envelope approximation, which is given by
$$i\left(\frac{\partial}{\partial
z}+\frac{1}{c}\frac{\partial}{\partial t}\right)\Omega_{p}+\frac{c}{2\omega_{p}}\left(\frac{\partial^2 }{\partial x^2}+\frac{\partial^2}{\partial y^2}\right)\Omega_{p}+\kappa_{13}\sigma_{31}=0, \eqno{(\textrm{S3})}$$
where $\kappa_{13}={\cal N}_a\omega_{p}|\mathbf{p}_{13}\cdot \hat{\mathbf{e}}_{p}|^2/(2\hbar\varepsilon_0 c)$ with ${\cal N}_a$ being atomic density and $c$ the vacuum's light speed. In the following calculations, the continuous-wave control field is assumed as strong enough so that $\Omega_c$ can be regarded as a constant during the evolution of the probe field.

The model under study can be realized in the realistic physical system, i.e., a ultracold $^{87}$Rb atomic gas. The energy levels can be selected from the underlying atoms tuned to D1-line transition, where $|1\rangle$, $|2\rangle$, and $|3\rangle$ are respectively set as $5^{2}\textrm{S}_{1/2}(F=1)$, $5^{2}\textrm{S}_{1/2}(F=2)$, and $5^{2}\textrm{P}_{1/2}(F=2)$ ~\cite{Steck}. Thus the system parameters are $\Gamma_2\simeq 1$ kHz, $\Gamma_3\simeq 5.75$ MHz, and $|\textbf{p}_{13}|=2.54\times10^{-27}$ C cm.
%
\section{Derivation of nonlinear envelope equation}{\label{Sec:2}}
In order to explore the nonlinear evolution of the probe field and the possible formation of the light gap bulets in the system, we derive nonlinear envelope equations of the probe field by employing the method of multiple scales based on the Maxwell-Bloch Eqs. (S2) and (S3). To this aim, the corresponding asymptotic expansion is arranged as the following form~\cite{HDP}: $\sigma_{jl}=\sum_{n=0}^{\infty}\epsilon^n\sigma_{jl}^{(m)}$, with $\sigma_{jl}^{(0)}=\delta_{j1}\delta_{l1}$,
$\Omega_{p}=\sum_{n=1}^{\infty}\epsilon^n \Omega_{p}^{(n)}$, $E_{s1}=\epsilon E_{s1}^{(1)}$, and $E_{s2}=\epsilon E_{s2}^{(1)}$. Thus it contributes to $d_{jl}=d_{jl}^{(0)}+\epsilon^{2}d_{jl}^{(2)}$, with
$d_{jl}^{(0)}=\Delta_j-\Delta_l+i\gamma_{jl}$ and $d_{jl}^{(2)}=
\frac{\alpha_{jl}}{2}[|E_{s1}^{(1)}|^2+|E_{s2}^{(1)}|^2]$. Here $\epsilon$ is the dimensionless small parameter expressing the typical amplitude of the probe field, and all the quantities on the right-hand side of the expansions are depended on the multi-scale variables $x_1=\epsilon x$, $y_1=\epsilon y$, $z_{n}=\epsilon^{n}z$ ($n=0,\,1,\,2$), and $t_{n}=\epsilon^{n}t$ ($n=0,\,1$). After substituting the above expansions into Maxwell-Bloch Eqs.~(S2) and~(S3) and comparing the coefficients of $\epsilon^n$, we can achieve a set of linear but inhomogeneous equations and then solve them order by order.

The first order ($n=1$) solution is given by $\Omega_{p}^{(1)}=\mathcal{F}e^{i\theta}$ and
$\sigma_{j1}^{(1)}=\{[\delta_{j3}(\omega+d_{21}^{(0)})-\delta_{j2}\Omega_c^\ast]/\mathcal{D}\}\mathcal{F}e^{i\theta}$,
where $\mathcal{D}=|\Omega_c|^2-(\omega+d_{21}^{(0)})(\omega+d_{31}^{(0)})$,
$\theta=\mathcal{K}(\omega)z_0-\omega t_0$, and $\mathcal{F}$ is a yet to be determined envelope function depending on the slow variables $x_1$, $y_1$, $z_1$, $z_2$, and $t_1$. According to the above solutions, we can obtain the linear dispersion relation: $\mathcal{K}(\omega)=\omega/c+\kappa_{13}(\omega+d_{21}^{(0)})/\mathcal{D}$. Tt is worth noting that the frequency and wave number of the probe field are $\omega_{p}+\omega$ and $k_{p}+\mathcal{K}(\omega)$, respectively. Thus $\omega=0$ corresponds to the center frequency of the probe field.

At the second order ($n=2$), we can get a solvability condition: $i[\partial /\partial z_{1}+(1/V_g)\partial/\partial t_1]\mathcal{F}=0$, with $V_g=(\partial \mathcal{K}/\partial \omega)^{-1}$ being the group velocity of the probe envelope.
The approximation solution at this order reads
$\sigma_{21}^{(2)}=a_{21}^{(2)}i\frac{\partial}{\partial t_1}\mathcal{F}e^{i\theta}$, $\sigma_{31}^{(2)}=a_{31}^{(2)}i\frac{\partial}{\partial t_1}\mathcal{F}e^{i\theta}$,
$\sigma_{jj}^{(2)}=a_{jj}^{(2)}|\mathcal{F}|^2e^{-2\bar{\alpha} z_2}$ ($j=\,1,\,2$), and
$\sigma_{32}^{(2)}=a_{32}^{(2)}|\mathcal{F}|^2e^{-2\bar{\alpha} z_2}$ with $\bar{\alpha}=\epsilon^{-2}{\rm Im}(\mathcal{K})$, where
%
$$a_{11}^{(2)}=\frac{[i\Gamma_{23}-2|\Omega_c|^2\mathcal{P}]\mathcal{G}-i\Gamma_{13}|\Omega_c|^2\mathcal{Q}}{i\Gamma_{13}|\Omega_c|^2\mathcal{P}^\ast}, \eqno{(\textrm{S4a})}\\$$
$$a_{22}^{(2)}=\frac{\mathcal{G}-i\Gamma_{13}a_{11}^{(2)}}{i\Gamma_{13}}, \eqno{(\textrm{S4b})}\\$$
$$a_{21}^{(2)}=-\frac{\Omega_c^\ast(2\omega+d_{21}^{(0)}+d_{31}^{(0)})}{\mathcal{D}^2}, \eqno{(\textrm{S4c})}\\$$
$$a_{31}^{(2)}=\frac{(\omega+d_{21}^{(0)})^2+|\Omega_c|^2}{\mathcal{D}^2}, \eqno{(\textrm{S4d})}\\$$
$$a_{32}^{(2)}=\frac{\Omega_c}{d_{32}^{(0)}}\left[\frac{1}{\mathcal{D}^\ast}-(a_{11}^{(2)}+2a_{22}^{(2)})\right], \eqno{(\textrm{S4e})}$$
%
with
$\mathcal{G}=(\omega+d_{21}^{(0)\ast})/\mathcal{D}^\ast-(\omega+d_{21}^{(0)})/\mathcal{D}$, $\mathcal{P}={1}/{d_{32}^{(0)}}-{1}/{d_{32}^{(0)\ast}}$, and $\mathcal{Q}={1}/({\mathcal{D}d_{32}^{(0)\ast}})-{1}/({\mathcal{D}^\ast d_{32}^{(0)}})$.

At the third order ($n=3$), a solvability condition yields the equation:
$$i\frac{\partial\mathcal{F}}{\partial z_2}-\frac{1}{2}\frac{\partial^2 \mathcal{K}}{\partial \omega^2}\frac{\partial^2 \mathcal{F}}{\partial t_1^2}+\frac{c}{2\omega_{p}}\left(\frac{\partial^2 }{\partial x_1^2}+\frac{\partial^2 }{\partial y_1^2}\right)\mathcal{F}+W_{1}|\mathcal{F}|^2\mathcal{F} e^{-2\bar{\alpha} z_2}+W_{2}V(x_1,y_1)\mathcal{F}=0. \eqno{(\textrm{S5})}$$
Here $W_1$ is the self-phase modulation coefficient, and $W_2$ is the cross-phase modulation coefficients contributed by the Stark fields. The explicit expressions of $W_{1}$, $W_{2}$, and $V(x_1,y_1)$ are respectively given by
%
$$W_{1}=\kappa_{13}\frac{\Omega_ca_{32}^{(2)\ast}+(\omega+d_{21}^{(0)})(2a_{11}^{(2)}+a_{22}^{(2)})}{\mathcal{D}}, \eqno{(\textrm{S6a})}\\$$
$$W_{2}=\kappa_{13}\frac{(\omega+d_{21}^{(0)})^2\alpha_{31}+|\Omega_c|^2\alpha_{21}}{2\mathcal{D}^2}, \eqno{(\textrm{S6b})}$$
$$V(x_1,y_1)=|E_{s1}^{(1)}(x_1)|^2+|E_{s2}^{(1)}(y_1)|^2. \eqno{(\textrm{S6c})}$$
%
Combining the solvability conditions (i.e., the equations for $\mathcal{F}$) at all orders, we obtain the unified equation for $\mathcal{F}$ in dimensionless form
%
$$i\frac{\partial \psi}{\partial s}+\frac{1}{2}\left(\frac{\partial^2 }{\partial \xi^2}+\frac{\partial^2 }{\partial \eta^2}-\beta\frac{\partial^2}{\partial \tau^2}\right)\psi+g|\psi|^2\psi+g^\prime V(\xi,\eta) \psi=-i\mathcal{A}\psi, \eqno{(\textrm{S7})}$$
with $\psi=\epsilon \mathcal{F}/\psi_0e^{-\bar{\alpha} z_2}$, $s=z/L_{\rm Diff}$, $\tau=[t-z/\textrm{Re}(V_g)]/\tau_0$, $(\xi,\eta)=(x,y)/R_\perp$, $\beta=L_{\rm Diff}/L_{\rm Disp}$, $g=L_{\rm Diff}/L_{\rm Nonl}$, $g^\prime=L_{\rm Diff}{\rm Re}(W_{2})E_{0}^2$, $V=[E_{s1}(\xi)^2+E_{s2}(\eta)^2]/E_{0}^2$, and $\mathcal{A}={\rm Im}(\mathcal{K}) L_{\rm Diff}$. Here $\psi_0$, $\tau_0$, $R_\perp$, and $E_{0}$ are respectively the typical Rabi frequency,  pulse duration,
beam radius, and field amplitude; $L_{\rm Diff}=\omega_pR_\perp^2/c$, $L_{\rm Disp}=\tau_0^2/\textrm{Re}(\partial^2 \mathcal{K}/\partial \omega^2)$, and $L_{\rm Nonl}=1/|{\rm Re}(W_{1})u_0^2|$ are the typical diffraction length, dispersion length, and nonlinear length, respectively.

Generally, the dimensionless coefficients in Eq.~(S7) are complex because the atomic system under consider is a lifetime broadened one and hence the system can not support stable light gap bullets. However, when the system works under the EIT circumstance, the imaginary parts of these coefficients can be minimized to extremely small values due to the quantum destructive interference effect~\cite{FIM}, which can be ignored safely in the following calculations.


For taking other physical parameters $\mathcal{N}_a\approx3.69\times10^{11}$ cm$^{-3}$, $\Omega_c=4.5\times10^{7}$ Hz, $\Delta_2=1.8\times 10^6$ Hz, and $\Delta_3=4.0\times 10^7$ Hz, the corresponding complex coefficients are given by $\mathcal{K}=(9.22+0.03i)$ cm$^{-1}$, ${\partial^2 \mathcal{K}}/{\partial \omega^2}=(2.37-0.18i)\times 10^{-13}$ cm$^{-1}$ s$^2$, $W_{1}=-(1.01+0.0045i)\times 10^{-14}$ cm$^{-1}$ s$^{2}$, $W_{2}=(3.26+0.021i)\times10^{-9}$ cm V$^{-1}$, where the imaginary parts of these quantities are indeed much smaller than their corresponding real parts. Furthermore, choosing $\psi_0=7.9\times10^{6}$ Hz, $\tau_0=6.16\times10^{-7}$s, $R=45$ $\mu$m, and $E_0=1.39\times10^{4}$ V cm$^{-1}$, thus all the three typical lengths ($L_{\rm Diff}$, $L_{\rm Disp}$, and $L_{\rm Nonl}$) are approximately equal, i.e., $L_{\rm Diff}=L_{\rm Disp}=L_{\rm Nonl}=1.59$ cm. As a result, we obtain $\beta=1-0.07i$, ${g}=-1-0.004i$, ${g}^\prime=1+0.006i$, and $\mathcal{A}=0.047$.
Assuming the form of the Stark field as $E_{s1}(\xi)=E_{s0}\cos \xi$ and $E_{s2}(\eta)=E_{s0}\cos \eta$ ($E_{s0}$ is a real constant), Eq.~(S7) is simplified as
$$i\frac{\partial \psi}{\partial s}=-\frac{1}{2}\left(\frac{\partial^2 \psi}{\partial \xi^2}+\frac{\partial^2 \psi}{\partial \eta^2}\right)+\frac{\beta}{2}\frac{\partial^2 \psi}{\partial \tau^2}+g|\psi|^2\psi+V_{\textrm{OL}}(\xi,\eta) \psi, \eqno{(\textrm{S8})}$$
%
where $V_{\textrm{OL}}(\xi,\eta)=-c_0(\cos^2 \xi+\cos^2 \eta$) and $\sqrt{c_0}=E_{s0}/E_{0}$.
